\documentclass[12pt]{article} 
\usepackage{epsf,latexsym}
\usepackage{amsmath,amssymb,amsthm,cite}
\usepackage{graphicx}

\theoremstyle{definition}                    

\theoremstyle{remark}

\numberwithin{equation}{section}             

\newcommand{\double}[1]{\mathbb{#1}}

\newcommand{\cc}{\double{C}}

\newcommand{\rr}{\double{R}}
\newcommand{\zz}{\double{Z}}

\newcommand{\kk}{\double{K}}

\newcommand{\aaa}{\mathcal{A}}

\newcommand{\hhh}{\double{H}}

\newcommand{\dd}{\mathcal{D}}

\newcommand{\op}{\oplus}

\newcommand{\bb}{\begin{eqnarray}}
\newcommand{\ee}{\end{eqnarray}}
\newcommand{\eee}{\nonumber\end{eqnarray}}

\usepackage[ps,dvips,arc,frame]{xy}

\newcommand{\rxy}[1]{{\begin{xy}
0;<2mm,0mm>:<0mm,2mm>::0;0,#1\end{xy}}}

\epsfverbosetrue
\newcommand{\T}{{\rm tr}}

\newcommand{\pp}[1]{\begin{pmatrix} #1 \end{pmatrix}}

\newcommand{\rxyz}[2]{{\begin{xy} 0;<2mm,0mm>:<0mm,2mm>::0;0,
,(5,-2)*{a}
,(10,-1.8)*{b}
,(15,-2)*{c}
,(20,-2)*{d}
,(25,-2)*{e}
,(30,-2)*{f}
,(2,-5)*{a}
,(2,-10)*{b}
,(2,-15)*{c}
,(2,-20)*{d}
,(2,-25)*{e}
,(2,-30)*{f}
,(5,-5)*\cir(#1,0){}
,(10,-5)*\cir(#1,0){}
,(15,-5)*\cir(#1,0){}
,(20,-5)*\cir(#1,0){}
,(25,-5)*\cir(#1,0){}
,(30,-5)*\cir(#1,0){}
,(5,-10)*\cir(#1,0){}
,(10,-10)*\cir(#1,0){}
,(15,-10)*\cir(#1,0){}
,(20,-10)*\cir(#1,0){}
,(25,-10)*\cir(#1,0){}
,(30,-10)*\cir(#1,0){}
,(5,-15)*\cir(#1,0){}
,(10,-15)*\cir(#1,0){}
,(15,-15)*\cir(#1,0){}
,(20,-15)*\cir(#1,0){}
,(25,-15)*\cir(#1,0){}
,(30,-15)*\cir(#1,0){}
,(5,-20)*\cir(#1,0){}
,(10,-20)*\cir(#1,0){}
,(15,-20)*\cir(#1,0){}
,(20,-20)*\cir(#1,0){}
,(25,-20)*\cir(#1,0){}
,(30,-20)*\cir(#1,0){}
,(5,-25)*\cir(#1,0){}
,(10,-25)*\cir(#1,0){}
,(15,-25)*\cir(#1,0){}
,(20,-25)*\cir(#1,0){}
,(25,-25)*\cir(#1,0){}
,(30,-25)*\cir(#1,0){}
,(5,-30)*\cir(#1,0){}
,(10,-30)*\cir(#1,0){}
,(15,-30)*\cir(#1,0){}
,(20,-30)*\cir(#1,0){}
,(25,-30)*\cir(#1,0){}
,(30,-30)*\cir(#1,0){}
#2\end{xy}}}

\newcommand{\rxyn}[2]{{\begin{xy} 0;<2mm,0mm>:<0mm,2mm>::0;0,
,(5,-2)*{a}
,(10,-1.8)*{b}
,(15,-2)*{c}
,(20,-2)*{d}
,(25,-2)*{e}
,(30,-2)*{f}
,(33,-2)*{\cdots}
,(2,-5)*{a}
,(2,-10)*{b}
,(2,-15)*{c}
,(2,-20)*{d}
,(2,-25)*{e}
,(2,-30)*{f}
,(2,-33)*{\vdots}
,(5,-5)*\cir(#1,0){}
,(10,-5)*\cir(#1,0){}
,(15,-5)*\cir(#1,0){}
,(20,-5)*\cir(#1,0){}
,(25,-5)*\cir(#1,0){}
,(30,-5)*\cir(#1,0){}
,(5,-10)*\cir(#1,0){}
,(10,-10)*\cir(#1,0){}
,(15,-10)*\cir(#1,0){}
,(20,-10)*\cir(#1,0){}
,(25,-10)*\cir(#1,0){}
,(30,-10)*\cir(#1,0){}
,(5,-15)*\cir(#1,0){}
,(10,-15)*\cir(#1,0){}
,(15,-15)*\cir(#1,0){}
,(20,-15)*\cir(#1,0){}
,(25,-15)*\cir(#1,0){}
,(30,-15)*\cir(#1,0){}
,(5,-20)*\cir(#1,0){}
,(10,-20)*\cir(#1,0){}
,(15,-20)*\cir(#1,0){}
,(20,-20)*\cir(#1,0){}
,(25,-20)*\cir(#1,0){}
,(30,-20)*\cir(#1,0){}
,(5,-25)*\cir(#1,0){}
,(10,-25)*\cir(#1,0){}
,(15,-25)*\cir(#1,0){}
,(20,-25)*\cir(#1,0){}
,(25,-25)*\cir(#1,0){}
,(30,-25)*\cir(#1,0){}
,(5,-30)*\cir(#1,0){}
,(10,-30)*\cir(#1,0){}
,(15,-30)*\cir(#1,0){}
,(20,-30)*\cir(#1,0){}
,(25,-30)*\cir(#1,0){}
,(30,-30)*\cir(#1,0){}
#2\end{xy}}}

\newcommand{\rxyg}[2]{{\begin{xy} 0;<2mm,0mm>:<0mm,2mm>::0;0,
,(5,-2)*{a} ,(10,-1.8)*{b} ,(15,-2)*{x_1} ,(20,-1.8)*{x_2} ,(2,-5)*{a}
,(1.8,-10)*{b} ,(2,-15)*{x_1} ,(1.8,-20)*{x_2} ,(5,-5)*\cir(#1,0){}
,(10,-5)*\cir(#1,0){} ,(15,-5)*\cir(#1,0){} ,(20,-5)*\cir(#1,0){}
,(5,-10)*\cir(#1,0){} ,(10,-10)*\cir(#1,0){} ,(15,-10)*\cir(#1,0){}
,(20,-10)*\cir(#1,0){} ,(5,-15)*\cir(#1,0){} ,(10,-15)*\cir(#1,0){}
,(15,-15)*\cir(#1,0){} ,(20,-15)*\cir(#1,0){} ,(5,-20)*\cir(#1,0){}
,(10,-20)*\cir(#1,0){} ,(15,-20)*\cir(#1,0){} ,(20,-20)*\cir(#1,0){}
#2\end{xy}}}

\newcommand{\rxym}[2]{{\begin{xy} 0;<2mm,0mm>:<0mm,2mm>::0;0,
,(5,-2)*{a}
,(10,-1.8)*{b}
,(15,-2)*{x_1}
,(20,-2)*{x_2}
,(25,-2)*{x_3}
,(30,-2)*{x_4}
,(33,-2)*{\cdots}
,(2,-5)*{a}
,(2,-10)*{b}
,(2,-15)*{x_1}
,(2,-20)*{x_2}
,(2,-25)*{x_3}
,(2,-30)*{x_4}
,(2,-33)*{\vdots}
,(5,-5)*\cir(#1,0){}
,(10,-5)*\cir(#1,0){}
,(15,-5)*\cir(#1,0){}
,(20,-5)*\cir(#1,0){}
,(25,-5)*\cir(#1,0){}
,(30,-5)*\cir(#1,0){}
,(5,-10)*\cir(#1,0){}
,(10,-10)*\cir(#1,0){}
,(15,-10)*\cir(#1,0){}
,(20,-10)*\cir(#1,0){}
,(25,-10)*\cir(#1,0){}
,(30,-10)*\cir(#1,0){}
,(5,-15)*\cir(#1,0){}
,(10,-15)*\cir(#1,0){}
,(15,-15)*\cir(#1,0){}
,(20,-15)*\cir(#1,0){}
,(25,-15)*\cir(#1,0){}
,(30,-15)*\cir(#1,0){}
,(5,-20)*\cir(#1,0){}
,(10,-20)*\cir(#1,0){}
,(15,-20)*\cir(#1,0){}
,(20,-20)*\cir(#1,0){}
,(25,-20)*\cir(#1,0){}
,(30,-20)*\cir(#1,0){}
,(5,-25)*\cir(#1,0){}
,(10,-25)*\cir(#1,0){}
,(15,-25)*\cir(#1,0){}
,(20,-25)*\cir(#1,0){}
,(25,-25)*\cir(#1,0){}
,(30,-25)*\cir(#1,0){}
,(5,-30)*\cir(#1,0){}
,(10,-30)*\cir(#1,0){}
,(15,-30)*\cir(#1,0){}
,(20,-30)*\cir(#1,0){}
,(25,-30)*\cir(#1,0){}
,(30,-30)*\cir(#1,0){}
#2\end{xy}}}

\begin{document}

\thispagestyle{empty}

\begin{center}

Institut f\"ur Mathematik  $^1$ \\ Universit\"at Potsdam
\\ Am Neuen Palais 10 \\14469 Potsdam \\ Germany\\

\vspace{2cm}

{\Large\textbf{Krajewski diagrams and the Standard Model}} \\

\vspace{1.5cm}

{\large Christoph A. Stephan $^2$}

\vspace{2cm}

{\large\textbf{Abstract}}
\end{center}

This paper  provides  a complete list of Krajewski diagrams 
representing  the standard model of particle physics. We will give
the possible representations of the algebra and the anomaly free lifts
which provide the representation of the standard model gauge group
on the fermionic Hilbert space. The algebra representations following
from the Krajewski diagrams are not
complete in the sense that the corresponding spectral triples 
do not necessarily obey to the
axiom of Poincar\'e duality. This defect may be repaired by
adding new particles to the model, i.e. by building models beyond
the  standard model.

The aim of this list of finite spectral triples (up to Poincar\'e duality)
is therefore to provide a basis for model building beyond the standard
model.

\vspace{2cm}

\noindent
PACS-92: 11.15 Gauge field theories\\
MSC-91: 81T13 Yang-Mills and other gauge theories

\vskip 1truecm

\noindent \\

\vspace{1.5cm}
\noindent $^1$ christophstephan@gmx.de\\

\newpage

\section{Introduction}

The aim of this paper is to provide a geometrical basis to enlarge
the standard model of particle physics within the setting of 
noncommutative geometry \cite{book,cc}. Krajewski diagrams \cite{kraj}
are a particularly useful tool to classify the finite part of spectral
triples \cite{kraj,pasch}. Especially the minimal Krajewski diagrams,
for finite spectral triples in $KO$-dimension zero as well as in $KO$-dimension
six, already allowed to reduce the number of interesting finite geometries
significantly \cite{class} and showed the singular role of the standard 
model within the class of almost-commutative geometries.

We will therefore give in this paper a complete list of minimal 
Krajewski diagrams which represent the finite part of spectral triples that allow 
to recover the first generation of the standard model in $KO$-dimension
zero as well as in the more promising $KO$-dimension six.
By {\it recovering the standard model} we mean that one
is able to reconstruct from the geometric data the
fermionic Hilbert space of the standard model with its well known gauge 
group $G_{SM}=U(1)_Y \times SU(2)_w \times SU(3)_c$ in the
correct representation and the correct charge assignment.

Only the first generation of the standard model without right-handed
neutrinos will be taken into account. The reason for this limitation
is the fact that noncommutative geometry does not explain why
there are three generations of Fermions.
Furthermore right-handed neutrinos usually appeared as a reducible
extension of the pure standard model \cite{class,mass1,DaweBar,barr}
and they can always be added later to each realisation of the standard
model shown below.

Depending whether one works in $KO$-dimension zero or six there are 
several mass mechanisms for the neutrinos available. In $KO$-dimension
zero they are usually based on the Higgs mechanism \cite{mass1,DaweBar}
while $KO$-dimension six allows also for Majorana masses \cite{barr,c06,mc2},
although at the expense of the orientability axiom \cite{ko6}. 
Recently A. Sitarz proposed   a third possibility \cite{sitarz} which builds
on a modification of the spectral action principle, resulting in
a radiative generation of neutrino masses. This mechanism is 
also compatible with all the models presented below.

The minimal Krajewski diagrams will in general respect all axioms
for spectral triples \cite{book}, save the axiom of the
Poincar\'e duality. For the first generation of the standard model,
restricted to the suitable finite matrix algebra \cite{cc,class}, 
the axiom holds of course.
But if the finite algebra is enlarged, new fermions are in general
needed \cite{AC,newcolour} to satisfy the Poincar\'e duality. 

Therefore the minimal Krajewski diagrams presented here
that do not necessarily satisfy the Poincar\'e duality, should 
serve as basic building
blocks to construct models beyond the standard model within
the noncommutative framework. They may also allow to push
the classification begun in \cite{class} further by enlarging 
the minimal diagrams in all possible ways. This has the advantage
that the standard model will always appear as a sub-model
and thus ensure the correct ``low energy limit'' of such
particle models. 

The paper is organised as follows: Starting with the basic
definitions of spectral triples and Krajewski diagrams we  fix the physical
requirements coming from the standard model and the resulting geometric
data. 
To obtain the correct hyper-charge assignment we will
use the central extension approach \cite{farewell}. The central
charges will then be fixed by the requirement of being free
of harmful anomalies, or equivalently by the requirement of
producing the standard model hyper-charge assignment.

Then we will construct the corresponding minimal Krajewski
diagrams in $KO$-dimension six and zero. We will start with
the more restrictive case of $KO$-dimension six and give in a 
second step the remaining diagrams in $KO$-dimension zero.

These basic minimal Krajewski diagrams can then be
used as building blocks for more sophisticated particle
models beyond the standard model.

\section{Basic Definitions}

In this section we will give the necessary basic definitions  for finite noncommutative
geometries \cite{book}. We will use the classical axioms and not the modified
versions of orientability and Poincar\'e duality as in \cite{mc2}.
We restrict ourselves to real, finite spectral triples
($\mathcal{A},\mathcal{H},\mathcal{D}, $ $J,\chi$). The algebra $\mathcal{A}$ is
a finite sum of matrix algebras
$\mathcal{A}= \oplus_{i=1}^{N} M_{n_i}(\mathbb{K}_i)$ with $\mathbb{K}_i=\mathbb{R},\mathbb{C},\mathbb{H}$ where $\mathbb{H}$
denotes the quaternions. 
A faithful representation $\rho$ of $\mathcal{A}$ is given on the finite dimensional Hilbert space $\mathcal{H}$.
The Dirac operator $\mathcal{D}$ is a selfadjoint operator on $\mathcal{H}$ and plays the role of the fermionic mass matrix.
$J$ is an antiunitary involution, $J^2=1$, and is interpreted as the charge conjugation
operator of particle physics.
The chirality $\chi$  is a unitary involution, $\chi^2=1$, whose eigenstates with eigenvalue
$+1$ are interpreted as right-handed particle states and left-handed antiparticle
states, whereas  the eigenstates with eigenvalue $-1$  represent the left-handed
particle states and right-handed antiparticle states.
These operators are required to fulfill Connes' axioms for spectral triples:

\begin{itemize}
\item  
$[J,\mathcal{D}]=\{J,\chi \}_{\pm}=0,$ $ \mathcal{D}\chi =-\chi \mathcal{D}$, 

$[\chi,\rho(a)]=[\rho(a),J\rho(a')J^{-1}]= [[\mathcal{D},\rho(a)],J\rho(a')J^{-1}]=0, \forall a,a' \in \mathcal{A}$,

where $\{J,\chi \}_{\pm}$ the commutator $[J,\chi]=0$ in $KO$-dimension zero 
and the anti-commutator $\{J,\chi \}=0$ in $KO$-dimension six.
\item The intersection form
$\cap_{ij}:=\T(\chi \,\rho (p_i) J \rho (p_j) J^{-1})$ is non-degenerate,
$\rm{det}\,\cap\not=0$. The
$p_i$ are minimal rank projections in $\mathcal{A}$. This condition is called
{\it Poincar\'e duality}. Demanding the Poincar\'e duality to hold requires
in $KO$-dimension six
an even number of summands in the matrix algebra \cite{barr,class}. 
\item The chirality can be written as a finite sum $\chi =\sum_i\rho(a_i)J\rho(a'_i)J^{-1}$,
which is a $0$-dim Hochschild cycle.
This condition is called {\it orientability}. 
\end{itemize} 

The representation $\rho ( a )$ takes the general  form
\bb
\rho ( a ) = \left( \oplus_{i,j=1}^N \rho(a_i,a_j) \right) \oplus  
\left(\oplus_{i,j=1}^N \overline{\rho^c}(a_i,a_j) \right)
\label{kranot}
\ee 
where $\rho(,)$ and $\rho^c(,)$ are the representation on
the particle and anti-particle Hilbert subspace. Without restricting
generality they can be taken to be
\bb
\rho(a_i,a_j) :=  a_i \otimes 1_{(m_{ij})}
\otimes 1_{(n_j)} \quad
\rho^c(a_i,a_j) := 1_{(n_i)} \otimes 1_{(m_{ij})} \otimes
a_j. 
\label{standard}
\ee
The multiplicities $(m_{ij})$ are non-negative integers. Here $(n)=n$ for $\kk=\rr,\cc$ and
$(n)=2n$ for $\kk=\hhh$.
We denote by
$1_(n)$ the $(n)\times (n)$ identity matrix and set by convention $1_0:=0$.
Algebra elements $a_i$ are taken to be from he $i$th summand 
$M_{n_i}(\kk_i)$ of the algebra $\aaa = \oplus_{i=1}^{N} M_{n_i}(\kk_i)$.

We will now present the basics of Krajewski diagrams, but 
only treat  the easy case, $\kk=\rr, \hhh$ in all components. For further
details on the complex case and on multiple arrows we 
refer to \cite{class}.
 
We define the {\it multiplicity matrix} $\mu \in M_N(\zz)$,
$N$ being the  number of summands in $\aaa$, such
that $\mu _{ij}:=\chi _{ij} \, m_{ij}$, with $m _{ij}$ being
the multiplicities of the representation (\ref{kranot}) and
$ \chi _{ij}$ the signs of the chirality. There are $N$ minimal projectors in
$\aaa$, each of the form
$p_i=0 \op  \cdots \oplus 0\op \rm{diag}(1_{(1)},0,...,0) \op 0\oplus \cdots
\op 0$. With respect to the basis $p_i$, the matrix of the
intersection form is $\cap = \mu \pm \mu ^T$, the relative plus (minus)
sign has its origin in the (anti-)commutation relation of the
real structure $J$ and the chirality $\chi$. 

If both entries $\mu _{ij}$ and $\mu _{ji}$ of the multiplicity
matrix are non-zero, then they must have the same (opposite) sign
in $KO$-dimension zero (six).

$\bullet$ Poincar\'e duality: The last condition to be satisfied by the
multiplicity matrix reflects the Poincar\'e duality
and requires the multiplicity matrix to obey  $\det(\cap =\mu \pm \mu ^T)\not=0$.
Since the intersection form is an anti-symmetric matrix  in $KO$-dimension six, 
this case restricts to an even number of summands in the matrix
algebra.

$\bullet$ The Dirac operator: The components of the (internal) Dirac
operator are represented by horizontal or vertical lines connecting two
nonvanishing entries of opposite signs in the multiplicity matrix $\mu $
and we will orient them from plus to minus. Each arrow represents a
nonvanishing, complex submatrix in the Dirac operator: For instance
$\mu_{ij}$ can be linked to $\mu_{ik}$ by
\begin{center}
\begin{tabular}{cc}
\rxy{
,(0,0)*\cir(0.7,0){}
,(5,0)*\cir(0.7,0){}
,(5,0);(0,0)**\dir{-}?(.6)*\dir{>}
,(0,-3)*{\mu_{ij}}
,(5,-3)*{\mu_{ik}}
}
\end{tabular} 
\end{center}
and this arrow represents respectively submatrices of $M$ in $\dd$ of
type $m\otimes 1_{(n_i)}$ with $m$ a complex $(n_j)\times(n_k)$ matrix.

\noindent Every arrow comes with three algebras:
Two algebras that localize its end
points, let us call them {\it right and left algebras}
and a third algebra that localizes the arrow, let us call it {\it colour
algebra}.  For the arrow presented above 
the left algebra is $\aaa _j$, the right algebra is $\aaa_k$ and the colour
algebra is $\aaa_i$.

We deduced  however in \cite{ko6} that if $i=j$ or $k=j$ 
the corresponding spectral triple
does not satisfy the axiom of orientability, so the colour algebra
must not coincide with the left of the right algebra. Translated into the language
of Krajewski diagrams this means that the arrow must not touch the
diagonal of the diagram. 

\noindent  We will restrict ourselves
to minimal Krajewski diagrams. A minimal Krajewski diagram
is defined in detail in \cite{algo},
in short it means that it is not possible to remove an arrow from the diagram
without changing the multiplicity matrix.

$\bullet$ Convention for the diagrams: 
Usually  arrows always point from right chirality for particles and 
antiparticles, to 
left chirality for particles and antiparticles. But since we will only
consider the general structure of the particle model and therefore
left-handedness and right-handedness are purely conventional,
we will not draw the arrowheads.
As a further convention the horizontal arrows will encode particles 
and its vertical copies  encode antiparticles. This choice is of course
also arbitrary. We will only draw the horizontal arrows in the Krajewski
diagrams below to keep them as uncluttered as possible.

\section{General requirements for the standard model}

To fix the geometrical data that will lead us to 
the minimal Krajewski diagrams, we assume as a physical
input only the first generation of the standard model without
right-handed neutrinos.

For the geometrical realisation there is a choice in
the so called $KO$-dimension of the spectral triple. In
physicists terms the $KO$-dimension can be thought
of as the signature of the metric of the internal space 
modulo 8. In this  sense $KO$-dimension six has the signature
$-2$, corresponding to the Minkowski version of the
finite spectral triple \cite{barr}. In the rather general construction 
presented below the $KO$-dimension is of little importance.
For $KO$-dimension six it only results in two extra constraints:
From the axiom of Poincar\'e duality follows that 
 the number of summands in the matrix algebra
has to be even. 
Also the representation of the algebra
is not allowed to represent the same summand on
a the same left- and right-handed particle species  and anti-particle 
species \cite{ko6}.

\subsection{The physical constraints}

As physical constraints we assume the following: 

\begin{itemize}
\item All standard model fermions, i.e. quarks and leptons, share for
their Dirac masses the same mass generating mechanism. 
This is the standard Higgs mechanism emerging from the spectral action
\cite{cc}.
\item We require the group of unitaries lifted to the Hilbert space
of the standard model fermions, to be the standard model gauge group
$G_{SM}=U(1)_Y \times SU(2)_w \times SU(3)_c$
\item We also require the models to be free of harmful anomalies, i.e. the
hyper-charge assignment is the one of the standard model.
\item For simplicity we will assume only one $U(1)$-subgroup in the standard
model gauge group, the hypercharge gauge group. 
It was shown in  \cite{farewell} that, due to the central extension, 
each additional $U(1)$-subgroup
results in an unphysical, completely decoupled extra photon. 
\end{itemize}

\subsection{The algebra and its representation}

Let us now construct the matrix algebra, its representation and
the internal Dirac operator which contains the Yukawa couplings.
Here we have to take care of the physical constraints specified
in the previous section as well as the axioms from noncommutative
geometry. 

From the standard model we know that the gauge group of 
any extension of the standard model 
has to contain $G_{SM}=U(1)_Y \times SU(2)_w \times SU(3)_c$
as a sub-group. In noncommutative geometry the non-abelian
part of the gauge group emerges as the group of unitary elements 
of the matrix algebra. This unitary group is then lifted to
the particle Hilbert space; we will cover the details of the lift in the next section.
For simplicity we choose as noncommutative subalgebra $\hhh \oplus M_3(\cc)$
which has as unitary group Aut$(\hhh \oplus M_3(\cc))=SU(2) 
\times U(3)$. But $M_2(\cc) \oplus M_3(\cc)$ will lead to similar results
with an extra $U(1)$ subgroup since Aut$M_2(\cc) \oplus M_3(\cc)=U(2) \times
U(3)$. This subtlety has no effect on the Krajewski diagrams, we will therefore
ignore it.

The abelian part of the gauge group emerges from a central extension 
of the lift, using the $U(1)$ subgroup of the $U(3)$ subgroup of the unitary
group.
To obtain the correct  $U(1)$ hyper-charge assignment of the standard model
the lift needs at least one abelian subalgebra $\cc$ of the matrix algebra
as a receptacle for the $U(1)$ group \cite{farewell}.  This leads directly to the minimal
matrix algebra $\aaa=\cc \oplus \hhh \oplus M_3(\cc)$ which is the 
valid candidate for the case of $KO$-dimension zero \cite{cc}. 
In $KO$-dimension six an even number of summands is needed and
one has to add a second copy of the complex numbers, i.e.
$\aaa=\cc \oplus \cc \oplus \hhh \oplus M_3(\cc)$ \cite{class}, if assuming the
classical axioms.

For the most general matrix algebra of a finite spectral triple containing the
standard model we find therefore
\bb
\aaa = M_3 (\cc) \oplus \hhh \oplus \bigoplus_{i=1}^{N-2} M_i(\kk) \ni (a,b,x_1,...,x_{N-2}),
\ee
with at least one summand being the complex numbers. We also assume
a finite number of summands with $N\geq 3$.

What is now the maximal number of summands equal to the complex
numbers which can affect the standard model particles? 
To determine this, we take the standard model with 
an algebra of four summands. Its Krajewski diagram is \cite{class}:
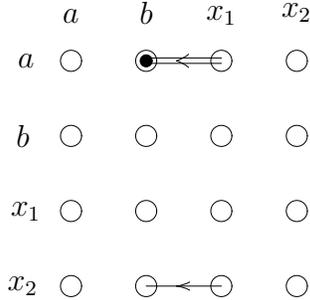
\begin{center}
\begin{tabular}{c}
\rxyg{0.7}{
,(10,-5)*\cir(0.4,0){}*\frm{*}
,(15,-5);(10,-5)**\dir2{-}?(.6)*\dir2{>}
,(15,-20);(10,-20)**\dir{-}?(.6)*\dir{>}
}
\\ \\
\end{tabular}
\end{center}
Here we have included the arrowheads in their standard form and
we have det$(\cap = \mu \pm \mu^t) \neq 0$ so the Poincar\'e duality is 
fulfilled. The algebra
of the model is
\bb
\aaa_{SM} = M_3 (\cc) \oplus \hhh \oplus \cc  \oplus \cc  \ni (a,b,x_1,x_2),
\label{sm1}
\ee
and its representation
\bb
\rho_{SM,L} (b) &=& \pp{ b \otimes 1_3 & 0 \\ 0 & b}, \quad \quad 
\rho_{SM,R} (x_1) = \pp{ x_1 1_3 & 0 & 0 \\ 0& \bar x_1 1_3 & 0 \\ 0& 0 & 
\bar x_1 1_2 },
\nonumber \\ \label{sm2} \\
\rho_{SM,L}^c (a,x_2)&=& \pp{1_2 \otimes a & 0 \\ 0 & x_2 1_2}, \quad  \quad 
\rho_{SM,R}^c (a,x_2) = \pp{a & 0 & 0 \\ 0 & a & 0 \\ 0& 0 & x_2 1_2}
\nonumber \\ \\
\rho_{SM} (a,b,x_1,x_2) &=& \rho_{SM,L} (b) \oplus \rho_{SM,R} (x_1) \oplus  
\overline{\rho_{SM,L}^c} (a,x_2) \oplus \overline{\rho_{SM,R}^c} (a,x_2).
\label{rep}
\ee

The Dirac operator takes the form
\bb
\mathcal{D} = \pp{\Delta & 0 \\ 0 & \bar{\Delta} },
\label{Dirac}
\ee
with the sub-matrices 
\bb
\Delta = \pp{0&0& M_d \otimes 1_3 & M_u \otimes 1_3 &0 \\
0 &0&0&0 & M_{e}\\
M_d^{\ast} \otimes 1_3 &0&0&0&0 \\
M_u^{\ast} \otimes 1_3 &0&0&0&0 \\
0&0&0&0&0 \\
0&M_{e}^{\ast} &0&0&0}
\label{Delta}
\ee
where $M_d$, $M_u$ and $M_e$ are in $M_{2\times 1}(\cc)$ and 
contain the Yukawa couplings
of the down-quark, the up-quark and the electron. 

The axioms of noncommutative geometry require now that 
$[\rho(a),J\rho(a')J^{-1}]= [[\mathcal{D},\rho(a)],J\rho(a')J^{-1}]=0$
which results in the following constraint: While the complex numbers
$x_2$
in the anti-particle representations 
$\rho_{SM,L}^c (a,x_2)$  and $\rho_{SM,R}^c (a,x_2)$
must have their origin in the
same summand of the matrix algebra this cannot be said for
three copies $x_1$ of the complex numbers in the particle representations
$\rho_{SM,R} (x_1)$. They
can, in principle, come from three different summands of complex
numbers. 

We conclude that we can accommodate at most four summands
of complex numbers in the matrix algebra which are represented on
the standard model fermions. Now the Krajewski diagram for
this model is
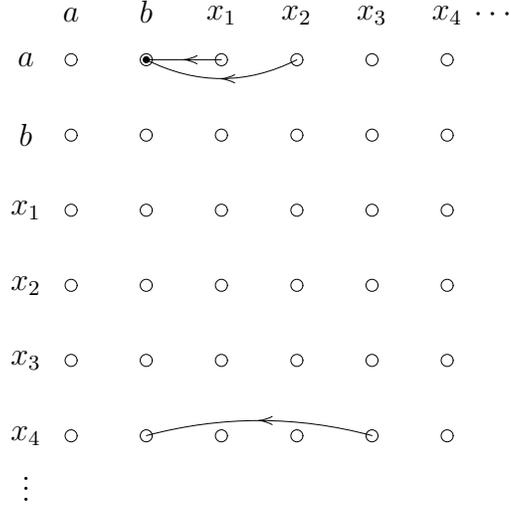
\begin{center}
\begin{tabular}{c}
\rxym{0.4}{
,(10,-5)*\cir(0.2,0){}*\frm{*}
,(10,-5);(15,-5)**\dir{-}?(.5)*\dir{<}
,(10,-5);(20,-5)**\crv{(15,-7.5)}?(.5)*\dir{<}
,(10,-30);(25,-30)**\crv{(17.5,-28)}?(.5)*\dir{<}
} \\ \\
\end{tabular}
\end{center}
where the dots indicate the more possible  summands in the matrix algebra.
If only the standard model fermions are included we find  
det$(\cap = \mu \pm \mu^t) = 0$ so the axiom of the Poincar\'e duality is 
not fulfilled. For a viable spectral triple more fermions, i.e. more arrows
have to be included. 

Ignoring the Poincar\'e duality for now, the matrix algebra has then the maximal
form
\bb
\aaa_{max} = M_3 (\cc) \oplus \hhh \oplus \cc_1 \oplus \cc_2 \oplus \cc_3 \oplus \cc_4
\oplus \bigoplus_{i=5}^{N-2} M_i(\kk) \ni (a,b,x_1,x_2,x_3,x_4,...,x_{N-2})
\label{max1}
\ee
where the first six summands are represented on the standard model
Hilbert subspace in the following way:
\bb
&&\rho_L (b) = \pp{ b \otimes 1_3 & 0 \\ 0 & b}, \quad \quad 
\rho_R (x_1,x_2,x_3) = \pp{ x_1 1_3 & 0 & 0 \\ 0& x_2 1_3 & 0 \\ 0& 0 & x_3 1_2 },
\nonumber \\ \label{max2} \\
&&\rho_L^c (a,x_4)= \pp{1_2 \otimes a & 0 \\ 0 & x_4 1_2}, \quad  \quad 
\rho_R^c (a,x_4) = \pp{a & 0 & 0 \\ 0 & a & 0 \\ 0& 0 & x_4 1_2} \ .
\nonumber
\ee
The Dirac operator for the standard model does not change. The
remaining part of the algebra $\bigoplus_{i=5}^{N-2} M_i(\kk)$,
its representation and corresponding part of the Dirac operator,
belong then to the ``beyond the standard model''  part and
have to be determined separately.

Smaller algebras with less summands 
represented on the standard model Hilbert space are readily obtained
by identifying two or more of the complex number summands. This
leads then, due to the necessary compatibility of the corresponding
representations, also to a representation of the standard
model.

\subsection{The lift and the Standard Model charges}

Let us now turn to the lift of the group of inner unitary group
of the matrix algebra $\aaa_{max}$. We will restrict ourselves
to the first six summands and their representation on the
Hilbert subspace of the standard model. The  group of unitaries is
$SU(2)  \times U(3)$ and contains
just a single $U(1)$ subgroup which is represented via
a central extension. 

The lift of the unitaries of $\aaa_{max}$ to 
the Hilbert space is in general given by $L=\rho J \rho J^{-1}$.
For the particle part of the standard model it takes the form
\bb
&&L^P \left( (\det u)^q \ u , v,  (\det u)^{p_1}, (\det u)^{p_2}, (\det u)^{p_3}, 
(\det u)^{p_4},... \right)|_{SM}
\nonumber \\ \\
&&= {\rm diag} [(\det u)^q\ v \otimes u, (\det u)^{p_4}\ v, (\det u)^{q+p_1}\ u,
(\det u)^{q+p_2}\ u, (\det u)^{p_3+p_4}],
\nonumber
\ee
where $v \in SU(2)$ and $u \in U(3)$. The central charges $p_i$ have
to be chosen to match the standard model representation of the
gauge group $G_{SM}=U(1)_Y \times SU(2)_w \times SU(3)_c$.

Comparing to the well known lift of the standard model \cite{farewell}
\bb
&&L^P_{SM} \left( (\det u)^q \ u , v,  (\det u)^{p}, (\det u)^{-p} \right)
\nonumber \\ \\
&&= {\rm diag} [(\det u)^q\ v \otimes u, (\det u)^{-p}\ v, (\det u)^{q+p}\ u,
(\det u)^{q-p}\ u, (\det u)^{-2p}]
\nonumber
\ee
with the relation
\bb
q = \frac{p-1}{3},
\ee
we find the following identifications that allow to recover the standard model
hyper-charge assignment:
\bb
p= p_1 = -p_2 = -p_3 = -p_4. 
\ee
It is now immediately clear why the $\cc$-summands in $\aaa_{max}$ may
be identified (if the axioms allow it): They all contribute the same central
charge, modulo a sign which can be obtained by taking the complex conjugate
in the respective representation.

\section{Implementing the constraints into the Krajewski diagrams}

We will now implement the physical constraints as well as the
constraints coming from the axioms into the Krajewski diagrams.
To keep the diagrams uncluttered we will only draw the arrows
representing the particles of the model. The  anti-particle arrows
are obtained by reflecting the particle arrows at the main diagonal.
All of the following constraints are therefore valid for the particle
arrows only but the anti-particles behave automatically in the correct
manner. 

We choose the first line and column of the diagram to represent
the $M_3(\cc)$ summand of the matrix algebra and the second
line and column the $\hhh$ summand. This already fixes the
double arrow of the quarks to lie on the first line with its connection
point at the second algebra, i.e. at the crossing of the first line
and the second column. Reading off the representation 
this would correspond to the particle part $\rho_L(b)=b \otimes 1_3$ and
the anti-particle part $\rho_L^c(a)=1_2 \otimes a$ (left-handedness and
right-handedness are again purely conventional). 

Since the colour of the quarks coming from the unitaries
of $M_3(\cc)$ is not broken by the standard model fermions, no particles
may connect to the first column \cite{class}. So arrows can only
connect on the second column and on columns further to the right
in the diagram. The choice of  a specific line to represent the 
``colour algebra'' of the quarks is of course also purely conventional. 

Also neither
quarks nor leptons couple vectorially to the $SU(2)$ subgroup and
therefore no arrows can lie on the second line of the diagram.
But both, leptons and quarks, couple with their right- or left-handed 
doublets chirally to the $SU(2)$ subgroup and therefore have to
connect to the second column. 

What left-handed or right-handed means is also conventional and
this choice is usually indicated by the direction of the arrow head.
To keep the diagrams here as general as possible, we will drop
the arrowheads.

The last physical constraint is that the leptons are neutral to 
the colour group. As a consequence the lepton arrow cannot 
lie on the $M_3(\cc)$-line, that is in our case the first line.

Putting these physical constraints together, we find the 
diagram depicted in figure 1, where a quark double arrow has been drawn to fix
the $M_3(\cc)$-line as well as the $\hhh$-line. 
Each line/column represents
a summand in the algebra $\aaa \ni (a,b,c,d,e,...)$ going from left to right. 
\begin{figure}
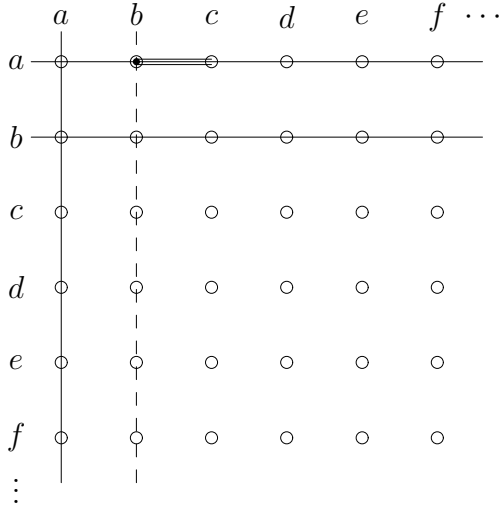

\begin{center}
\begin{tabular}{c}
\rxyn{0.4}{
,(10,-5)*\cir(0.2,0){}*\frm{*}
,(10,-5);(15,-5)**\dir2{-}
,(3,-5);(33,-5)**\dir{-}
,(3,-10);(33,-10)**\dir{-}
,(5,-3);(5,-33)**\dir{-}
,(10,-3);(10,-33)**\dir{--}
}
\end{tabular}
\caption{Diagram for $KO$-dimension zero with a quark double arrow drawn in. 
The dashed column
represents the $\hhh$-line to which $SU(2)$ doublets have to connect. 
The continuous lines and columns are prohibited for the lepton arrow.}
\end{center}
\end{figure}
The connected end of the quark arrow represents the $SU(2)$ doublet
and the two ends the $U(1)$ singlets.  Continuous lines 
represent the physical constraints specified above.
The lepton arrow will be added next has to connect to the 
second column, accentuated by the dashed line, and must not
connect to or lie on any of the continuous lines.  

In the case of $KO$-dimension six we have an additional constraint from
the orientability axiom \cite{ko6}. It translates into the requirement
that no arrow, including the quark arrow, 
may connect to the main diagonal. We depict
this by another continuous line in figure 2.
\begin{figure}
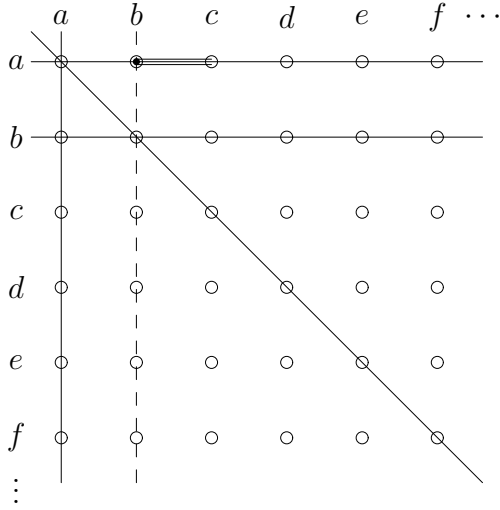

\begin{center}
\begin{tabular}{c}
\rxyn{0.4}{
,(10,-5)*\cir(0.2,0){}*\frm{*}
,(10,-5);(15,-5)**\dir2{-}
,(3,-5);(33,-5)**\dir{-}
,(3,-10);(33,-10)**\dir{-}
,(5,-3);(5,-33)**\dir{-}
,(10,-3);(10,-33)**\dir{--}
,(3,-3);(33,-33)**\dir{-}
}
\end{tabular}
\caption{Diagram for $KO$-dim six with a quark double arrow drawn in. 
The dashed column
represents the $\hhh$-column to which $SU(2)$ doublets have to connect. 
The continuous lines and columns are prohibited for the lepton arrow.
No arrow is allowed to connect to the continuously drawn diagonal.}
\end{center}
\end{figure}

\subsection{The Krajewski diagrams of the Standard Model}

To construct the full list of Krajewski diagrams of the standard model
(up to Poincar\'e duality) we start as before with the quarks to fix the 
first three or four summands of the matrix algebra.  

We note that the spectral triples are invariant under simultaneous permutations
of lines and columns of the respective Krajewski diagrams. These permutations
result only in a reshuffle of the algebra's summands, its representation, Hilbert
space and corresponding Dirac operator. But they do not alter the physical
content of the theory \cite{algo}. Therefore Krajewski diagrams which can
be obtained by permutations are equivalent.

Figure 3 shows the two possible ways to put the quarks into a Krajewski diagram. 
The algebra truncated to three summands of the left diagram is 
$\aaa=M_3(\cc) \oplus \hhh \oplus \cc \ni (a,b,c)$ with the representation
$\rho_L(b) = b \otimes 1_3$, $\rho_R(c) =$diag$(c  1_3, \bar{c} 1_3)$,
$\rho^c_L(a) =1_2 \otimes a$ and $\rho^c_R(a) =$diag$(a,a)$. For the left diagram in figure
3 we have $\aaa=M_3(\cc) \oplus \hhh \oplus \cc  \oplus \cc \ni (a,b,c,d)$ with the representation
$\rho_L(b) = b \otimes 1_3$, $\rho_R(c,d) =$diag$(c  1_3, d 1_3)$,
$\rho^c_L(a) =1_2 \otimes a$ and $\rho^c_R(a) =$diag$(a,a)$.
\begin{figure}
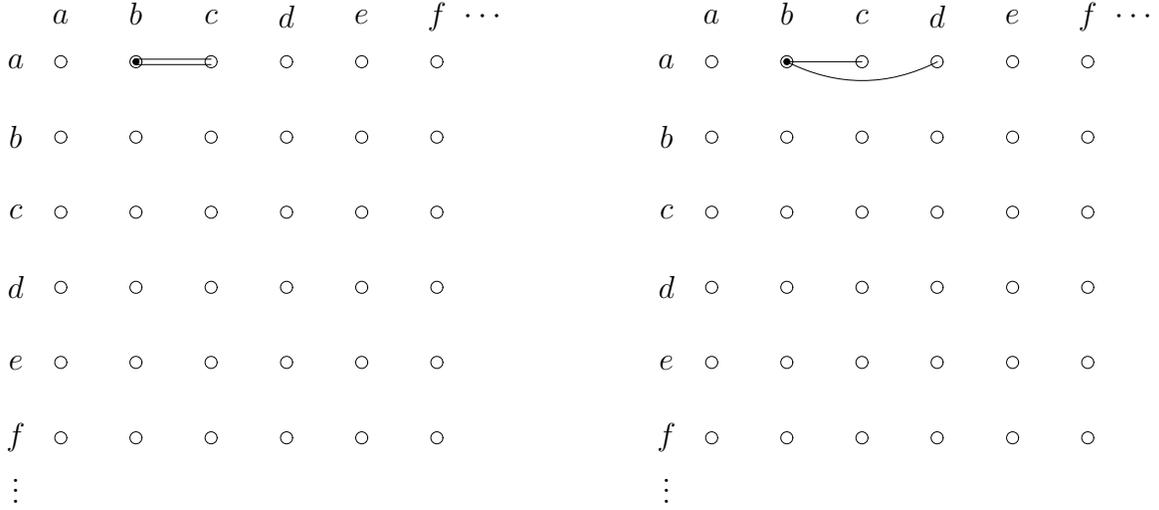

\begin{center}
\begin{tabular}{ccc}
\rxyn{0.4}{
,(10,-5)*\cir(0.2,0){}*\frm{*}
,(10,-5);(15,-5)**\dir2{-}
}
&
\quad \quad \quad \quad
&
\rxyn{0.4}{
,(10,-5)*\cir(0.2,0){}*\frm{*}
,(10,-5);(15,-5)**\dir{-}
,(10,-5);(20,-5)**\crv{(15,-7.5)}
}

\end{tabular}
\caption{The two representation of the standard model quarks in  a Krajewski
diagram. All other possibilities can be obtained by simultaneous permutation
of the lines and the columns.}
\end{center}
\end{figure}

Let us now add the lepton arrow according to the physical and
geometrical constraints depicted in figure 1 and figure 2.
We will begin with the more restrictive case of a finite spectral 
triple in $KO$-dimension six as shown in figure 2. 

Building on the left diagram of figure 3 we add a lepton arrow on
the third line, the first allowed line. Connecting it according to the
rules to the second column, the closest end point is at the third 
column. The whole diagram is shown in figure 4 together with
a possible permutation obtained by interchanging the fourth and
fifth line/column ($d \leftrightarrow e$).  
The algebra and its  representation truncated to the standard model
are \\ $\aaa=M_3(\cc) \oplus \hhh \oplus \cc \oplus \cc \ni (a,b,c,d)$
with 
\bb
&&\rho_L (b) = \pp{ b \otimes 1_3 & 0 \\ 0 & b}, \quad \quad 
\rho_R (c,d) = \pp{ c 1_3 & 0 & 0 \\ 0& \bar{c} 1_3 & 0 \\ 0& 0 & d 1_2 },
\nonumber \\ \\
&&\rho_L^c (a,c)= \pp{1_2 \otimes a & 0 \\ 0 & \bar{c} 1_2}, \quad  \quad 
\rho_R^c (a,c) = \pp{a & 0 & 0 \\ 0 & a & 0 \\ 0& 0 & \bar{c} 1_2} \ .
\nonumber
\ee
The Dirac operator and the other operators appearing in the spectral
triple remain the same for all realisations of the standard model.

The next in-equivalent diagram is shown in figure 5 together with an
equivalent diagram obtained by permuting the fourth and fifth line/column
($d \leftrightarrow e$). Note that the  left diagram in figure 5, when 
truncated to the first four summands in  the algebra, corresponds
exactly to the Krajewski diagram of the  
minimal standard model found in the classification \cite{class}.
Its algebra and representation are given by (\ref{sm1}) and (\ref{sm2})
with the identification $c = x_1$ and $d=x_2$.
For the following diagrams we will not give the details of the algebra
and its representation.

We proceed in this spirit  (only depicting one representative for 
each equivalence class of Krajewski diagrams) and find five more
diagrams which concur with the physical and geometrical constraints for 
finite spectral triples with $KO$-dimension six. These five diagrams
are shown in figures 6, 7 and 8.

In $KO$-dimension zero, the conditions on the Krajewski diagrams are more
relaxed since the lepton arrow may touch the diagonal, see figure 1.
The previous seven in-equivalent diagrams shown in figures 3-8 
are also admissible in $KO$-dimension zero but we find four more
diagrams, see figure 9 and figure 10. Note again that the
left diagram in figure  9, if truncated to the  first three summands
in the algebra, is the Krajewski diagram \cite{kraj} which 
represents the classical version of the noncommutative
standard model by A. Chamseddine and A. Connes \cite{cc}.

\section{Conclusions}

In this paper we have presented all in-equivalent Krajewski diagrams
which represent spectral triples constituting the first family of the
standard model of particle physics without right-handed neutrinos.
We have ignored for  the  moment the axiom of Poincar\'e
duality \cite{book}, which is of course respected for suitable
truncations leading to the well known Krajewski diagrams of the
standard model, i.e. figure 5 and figure 9 (left) truncated at four summands
or three summands.

We find eleven in-equivalent diagrams for spectral triples with
$KO$-dimension zero. Of these eleven diagrams the first seven, figures 4-8
are also compatible with the more restrictive conditions for
spectral triples with $KO$-dim six.

The eleven Krajewski diagrams may now be used as basic building blocks 
for models beyond the standard model. Only a few models
beyond the standard model are known within noncommutative
geometry \cite{AC,newcolour,vector} and these extensions
have been found by trial and error methods. 
Now it appears to be possible to explore the realm beyond the standard
model in a more organised way by starting with one of the standard model
diagrams presented here and extending it by enlarging the number
of summands in the algebra and its particle content. Thereby 
one is always sure to obtain the standard model as a sub-model.

This procedure will still be extremely restricted, not only by the
axiom of Poincar\'e duality that should be obeyed by the final model.
But also the spectral action principle poses extra constraints
on the physical models \cite{cc,mc2} which result for example
in restrictions on the masses and gauge couplings 
of the new particles as it is the case for the $\theta$-particle
model, \cite{newcolour}. As an example let us provide
its  Krajewski diagram (with the arrowheads put into place)
which consists of an extension of diagram 5:
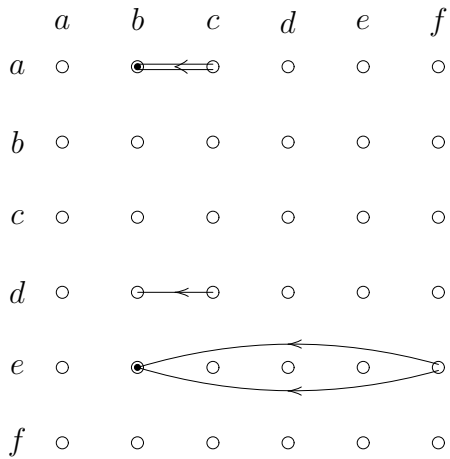
\begin{center}
\begin{tabular}{c}
\rxyz{0.4}{
,(10,-5)*\cir(0.2,0){}*\frm{*}
,(10,-5);(15,-5)**\dir2{-}?(.5)*\dir2{<}
,(10,-20);(15,-20)**\dir{-}?(.5)*\dir{<}
,(10,-25);(30,-24.8)**\crv{(20,-22)}?(.5)*\dir{<}
,(10,-25);(30,-25.2)**\crv{(20,-28)}?(.5)*\dir{<}
,(10,-25)*\cir(0.2,0){}*\frm{*}
} \\ \\
\end{tabular}
\end{center}
Since we know that extensions of the standard model within the
noncommutative framework lead to models of physical interest like
the AC-model \cite{AC}, which even provides an interesting dark 
matter candidate \cite{klop}, this endeavour to seek for new
physics seems very promising.

\subsection*{Acknowledgements}

The author gratefully acknowledges the funding of his work
by the Deutsche Forschungsgemeinschaft.

\begin{figure}[h]
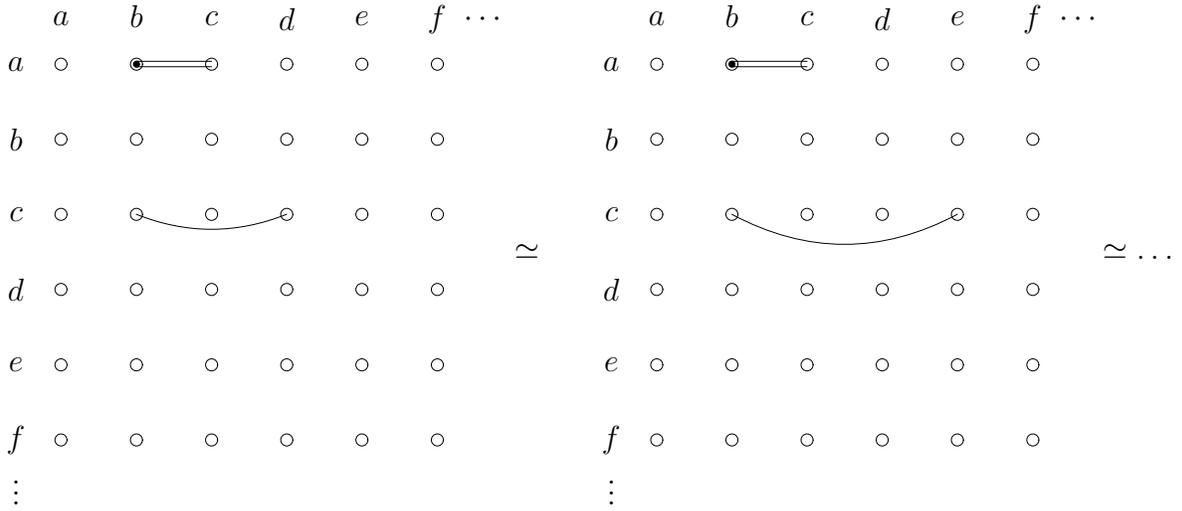

\begin{center}
\begin{tabular}{cc}
\rxyn{0.4}{
,(10,-5)*\cir(0.2,0){}*\frm{*}
,(10,-5);(15,-5)**\dir2{-}
,(10,-15);(20,-15)**\crv{(15,-17)}
,(37,-17.5)*{\simeq \quad}
}
&
\rxyn{0.4}{
,(10,-5)*\cir(0.2,0){}*\frm{*}
,(10,-5);(15,-5)**\dir2{-}
,(10,-15);(25,-15)**\crv{(17.5,-19)}
,(37,-17.5)*{\simeq \dots}
}
\end{tabular}
\caption{Krajewski diagram with one quark double arrow (left diagram in figure 3)
and one lepton arrow according to the restrictions for $KO$-dimension six. The two
diagrams show two possible permutation, i.e. $d \leftrightarrow e$, giving
equivalent diagrams.}
\end{center}
\end{figure}

\begin{figure}[h]
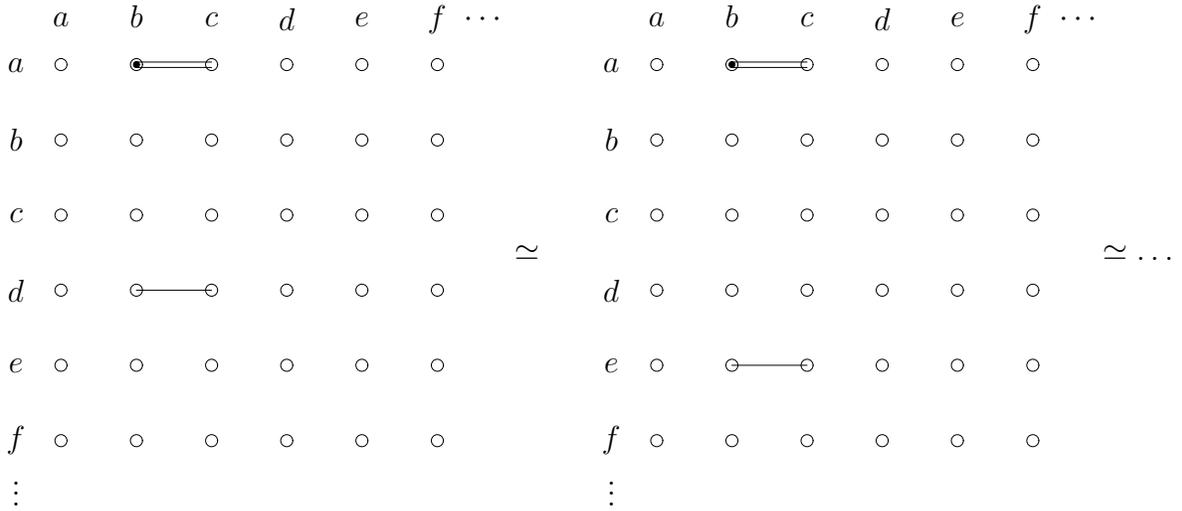

\begin{center}
\begin{tabular}{cc}
\rxyn{0.4}{
,(10,-5)*\cir(0.2,0){}*\frm{*}
,(10,-5);(15,-5)**\dir2{-}
,(10,-20);(15,-20)**\dir{-}
,(37,-17.5)*{\simeq \quad}
}
&
\rxyn{0.4}{
,(10,-5)*\cir(0.2,0){}*\frm{*}
,(10,-5);(15,-5)**\dir2{-}
,(10,-25);(15,-25)**\dir{-}
,(37,-17.5)*{\simeq \dots}
}
\end{tabular}
\caption{Krajewski diagram of the  standard model with constraints compatible
with $KO$-dimension six. This diagram is in-equivalent to the Krajewski diagram
shown in figure 4. The permutation $d \leftrightarrow e$ leads to the equivalent
diagram on the right.}
\end{center}
\end{figure}

\begin{figure}[h!]
\begin{center}
\begin{tabular}{ccc}
\rxyn{0.4}{
,(10,-5)*\cir(0.2,0){}*\frm{*}
,(10,-5);(15,-5)**\dir2{-}
,(10,-20);(25,-20)**\crv{(17.5,-23)}
}
& \quad \quad \quad \quad &
\rxyn{0.4}{
,(10,-5)*\cir(0.2,0){}*\frm{*}
,(10,-5);(15,-5)**\dir{-}
,(10,-5);(20,-5)**\crv{(15,-7.5)}
,(10,-15);(20,-15)**\crv{(15,-17)}
}
\end{tabular}
\caption{In-equivalent Krajewski diagrams compatible with $KO$-dimension six.}
\end{center}
\end{figure}

\begin{figure}[h!]
\begin{center}
\begin{tabular}{ccc}
\rxyn{0.4}{
,(10,-5)*\cir(0.2,0){}*\frm{*}
,(10,-5);(15,-5)**\dir{-}
,(10,-5);(20,-5)**\crv{(15,-7.5)}
,(10,-15);(25,-15)**\crv{(17.5,-18)}
}
& \quad \quad \quad \quad &
\rxyn{0.4}{
,(10,-5)*\cir(0.2,0){}*\frm{*}
,(10,-5);(15,-5)**\dir{-}
,(10,-5);(20,-5)**\crv{(15,-7.5)}
,(10,-25);(20,-25)**\crv{(15,-27)}
}
\end{tabular}
\caption{In-equivalent Krajewski diagrams compatible with $KO$-dimension six.}
\end{center}
\end{figure}

\begin{figure}[h!]
\begin{center}
\begin{tabular}{c}
\rxyn{0.4}{
,(10,-5)*\cir(0.2,0){}*\frm{*}
,(10,-5);(15,-5)**\dir{-}
,(10,-5);(20,-5)**\crv{(15,-7.5)}
,(10,-25);(30,-25)**\crv{(20,-28)}
}
\end{tabular}
\caption{In-equivalent Krajewski diagram compatible with $KO$-dimension six.}
\end{center}
\end{figure}

\begin{figure}[h!]
\begin{center}
\begin{tabular}{ccc}
\rxyn{0.4}{
,(10,-5)*\cir(0.2,0){}*\frm{*}
,(10,-5);(15,-5)**\dir2{-}
,(10,-15);(15,-15)**\dir{-}
}
& \quad \quad \quad \quad &
\rxyn{0.4}{
,(10,-5)*\cir(0.2,0){}*\frm{*}
,(10,-5);(15,-5)**\dir2{-}
,(10,-20);(20,-20)**\crv{(15,-23)}
}
\end{tabular}
\caption{In-equivalent Krajewski diagram compatible with $KO$-dimension zero.}
\end{center}
\end{figure}

\begin{figure}[h!]
\begin{center}
\begin{tabular}{ccc}
\rxyn{0.4}{
,(10,-5)*\cir(0.2,0){}*\frm{*}
,(10,-5);(15,-5)**\dir{-}
,(10,-5);(20,-5)**\crv{(15,-7.5)}
,(10,-15);(15,-15)**\dir{-}
}
& \quad \quad \quad \quad &
\rxyn{0.4}{
,(10,-5)*\cir(0.2,0){}*\frm{*}
,(10,-5);(15,-5)**\dir{-}
,(10,-5);(20,-5)**\crv{(15,-7.5)}
,(10,-25);(25,-25)**\crv{(17.5,-28)}
}
\end{tabular}
\caption{In-equivalent Krajewski diagram compatible with $KO$-dimension zero.}
\end{center}
\end{figure}

\end{document}